# LA City Bike Lane Infrastructure and its Effects on Business Closures From 2016-2021

Jeff Block, Yocelyne Gomez, Carolina Gonzalez, Miguel Magana, Marco Rodas
**Academic Advisor and Mentor: Dr. Jongwook Woo**

FEMBA, California State University Los Angeles
{jblock, ygomez5, cgonz264, mmagana51, mrodas13, **jwoo5**}@calstatela.edu

**Abstract:** This paper analyzes the presence of bike lanes within the city of Los Angeles and how their presence could offer positive economic benefits to businesses. City wide shutdowns initiated in March of 2020 acted as a demand shock to local businesses. Over the next two years, businesses struggled to keep the doors open. We explored the possibility of bicycle infrastructure playing a role in supporting/insulating local businesses from closure and positive economic impacts. Our analysis shows that the businesses along bike lane in Los Angeles runs positively, and that COVID-19 is not related to, significantly.

## 1. Introduction

We use SAP Analytics Cloud (SAC) to process and visualize the historical data of bike lanes through the city of LA between 2016 and 2021. With gas prices rising to over $6 as of March 2022, residents of LA look to alternative means of transportation [3]. Thankfully, bicycle advocacy groups have long pushed for the expansion of bike lane infrastructure. Unfortunately, bike lane expansion often comes at the expense of local business parking and increases in traffic [4]. To understand economic and business impacts of bike lanes on businesses we look to the data provided by the city of Los Angeles. Combining multiple data sets into relevant visualizations we are then able to make judgments on the significance of bike lanes.

We look at two sections of road spanning a continuous 1.7 miles of Los Angeles, Sunset Blvd. in Silverlake, CA (Glendale Blvd to Santa Monica Blvd) will be our **BIKE LANE** area of study. Santa Monica Blvd in Hollywood, CA (Hoover Blvd to Van Ness Blvd) will be our **NON-BIKE LANE** area of study. The two areas of study flow directly into one another forming a continuous road with similar local business and approximately the same demographic composition [5]. Both areas will look at business within a 500' setback from the bike lane/non-bike lane.

## 2. Related Work

It has long been argued that bike lanes provide a positive economic impact on cities in addition to providing alternative/low cost means of transportation. Findings by Transportation Research and Education refute assertions made by bike infrastructure critics that new lanes hurt adjacent businesses by making car access less convenient [2]. The question returns to what is the measurable economic impact of bike lanes and how that impact might help make a case for cities to invest in the expansion of lane infrastructure.

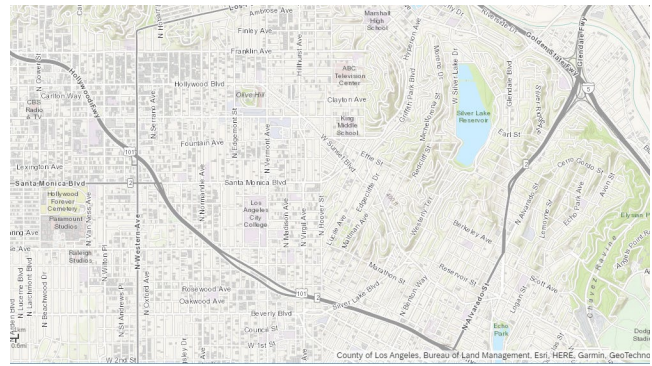

Figure 1. Map of area of study

We assume that the widespread closures due to the COVID-19 pandemic lockdowns acted as a demand shock to local businesses. A recent study of more than 5,800 small businesses shed light on the economic impact of COVID-19 shutdowns illustrating an increased level of financial fragility in the weeks after the COVID-19 related shutdowns [6].

Our paper attempt to connect the presence of bike lanes to increased levels of financial and economic resilience on the part of local businesses.

## 3. Specifications

The data was retrieved from lacity.org [7, 8]. The website is an open source collection of data sets meant to be used to explore information about the City of Los Angeles. The data is updated regularly by Mayor Eric Garcetti's Data Team as part of the official Medium Publication. Out of the 1,622 sets available, the 3 used are listed in Table 1 below.

The Bike Lanes data was last updated in November 2020. Given that the physical city streets don't change often, the data can be considered current. The business data is updated monthly, last updated March 15, 2022.

The business data for active and closed business were 67.1MB and 47.5MB in size, respectively. District 13 was the focus of this paper for the years 2016 to 2021. The data contained location information by latitude and longitude. With that detail, the businesses were plotted on the bike lane map for the area of interest. The extracted data totaled 6.7MB as detailed in Table 1.

Table 1. Data Specifications - Bike Lanes and Businesses

| Data Set | Size (Total 6,663 KB) |
|---|---|
| Listing of Active Businesses (Office of Finance) | 3,706 KB |
| Bike Lanes (LA DOT) | 123 KB |
| All Closed Businesses (Office of Finance) | 2,834 KB |

## 4. Implementation Flowchart

The RAW dataset consisted of two downloads from the LA City Websites, the first comprised the location of all Bike Lanes within the city. The second consisted of the complete LA City Business Registration dataset, including active businesses and closed businesses in the map area of study. The datasets were cleaned to reflect information needed for our analysis. The entire data manipulation process is shown in the below flowchart (Figure 2). In addition, there are 5 data logs in CSV format uploaded to SAP. After the upload was completed, we created the Models and Stories, which provided our visualizations in Figures 5, 6, and 7. The Stories and Models will be exported and formatted for use in PowerPoint.

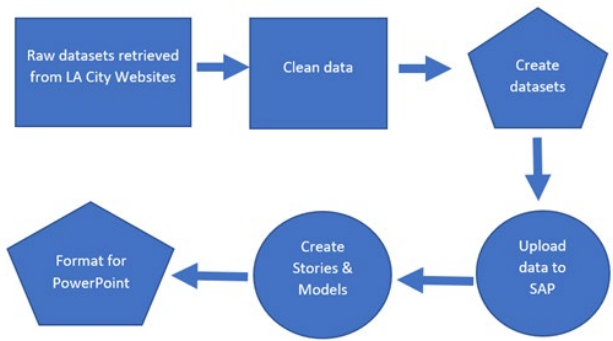

Figure 2. Implementation Flowchart

## 5. Data Cleaning

Before the CSV file could be uploaded onto SAP Analytics cloud as a data sheet it first had to be cleaned of excess information not relevant to the project. Taking the Listing of Active Businesses CSV file, the first step restricts the information to only the years 2016-2021. This information was copied over to a new file for extraction. Further cleaning the data, we extracted the latitude and longitude. Missing and blank locations were eliminated and the remaining data was combined with the zip+4 information selecting for LA District 13 (Silverlake/West Hollywood).

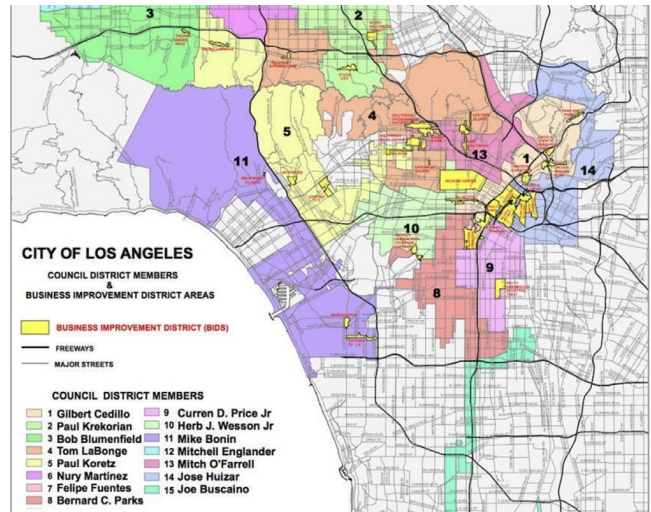

Figure 3. City of LA district MAP (district 13)

The remaining data was selected for business status (20 for Active and 30 for Closed). With the new clean data of businesses either open or closed (plus lat/long) we were able to upload the data set into SAP. This data set was then combined with the LA City Bike Lane (LA DOT).

## 6. Analysis and Visualization

After data cleaning and determining what data would best describe our analysis, a story and predictive model were created in SAP Analytics Cloud that provided a visual representation of the location and density of business closures within the city of Los Angeles.

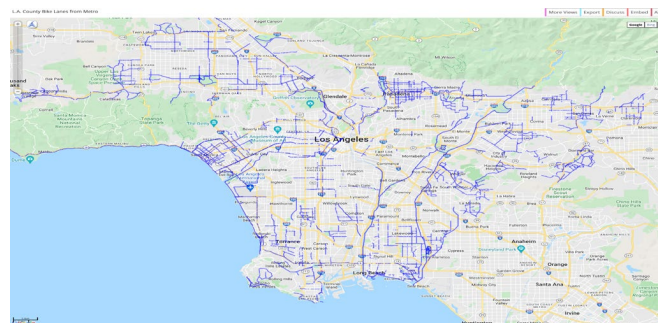

Figure 4. LA City Bike Lane network.

### 6.1 Closed and Open Bar Chart

Before taking an in-depth review of the businesses that opened and closed from 2016-202, a bar chart provides us with a visualization of the fluctuation of businesses opening and closing over the timeline. This bar chart depicts business activity by quarters and at first glance trends are particularly apparent. New businesses were more likely to open within the first quarter of the year based on volume. The opposite was true businesses that closed which happened primarily in the last quarter of the year.

By adjusting the "Drill" levels in the chart the data displayed in the bar chart can reflect business activity by year. From 2016 to 2018 the average new businesses annually were 2,151 and closed businesses were 2,087. From 2019 to 2021 the new annual businesses opened dropped to 1,804, but was more significant for 2021 was the amount of businesses that closed which dropped to 506 or 32% drop from the previous year.

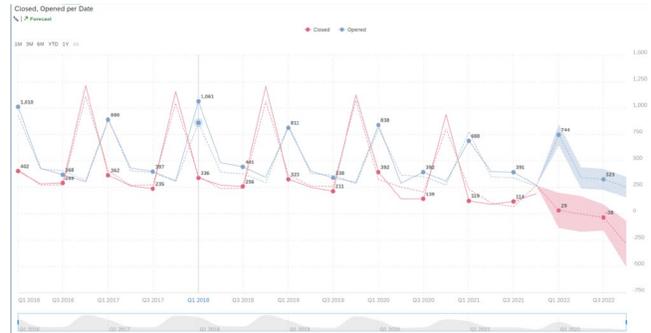

Figure 6. Open/Closed Time Series with Forecast

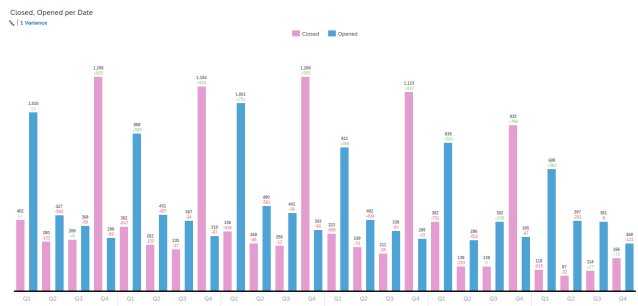

Figure 5. 2016-2021 Open/Closed bar chart

### 6.2 Time Series Chart with Forecast

The Time Series chart provides an additional visualization of business activity in District 13. It displays a gradual downtrend from 2016 through 2021 for both new and closed business. The forecast function with triple exponential smoothing was applied to the Time Series Chart. While the chart below reflects significant drops in new and closed businesses projected for 2022, the Time Series Chart can be adjusted by focusing and by adjusting the visualization in the chart to focus only on 2021 and the forecast; it appears to be more consistent with the performance in 2021.

### 6.3 Heat Map

To assist with determining which area to focus on within District 13, the bike lanes are clearly identified, buy utilizing a heat map. It can then be determined which areas have higher concentrations of closed businesses. The heat map identified two potential stretches of roads to compare. The bike lane route towards Hollywood Blvd. and Santa Monica Blvd. both had concentrations of closed businesses, however Santa Monica Blvd. was a closer reflection of the businesses on the bike lane. Hollywood Blvd has a very dense concentration of businesses and the closure activity (yellow) is reflective of that. Santa Monica Blvd. was selected as the "Non-Bike Lane" road.

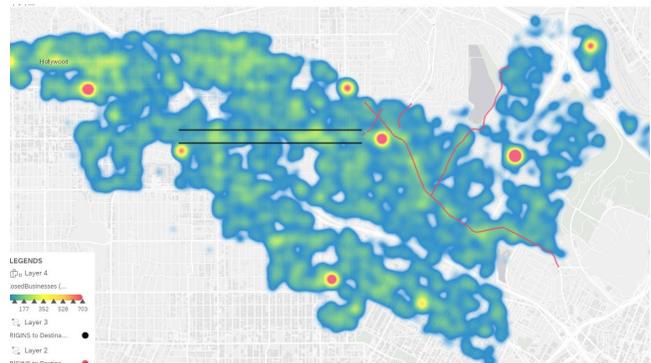

Figure 7. Area of study Geo Map with open/closed.

### 6.4 Geo Map

The focus of our study was Council District 13 and a Geo Map was used to provide visual reference of our focus and visualization for the paper. Businesses that opened and closed from 2016-2021 are represented by blue squares and red stars throughout the map. Additional parameters added to the Geo Map were flow layers which represented the bike lanes in District 13 as well as the non-bike lane area. With the flow layers activated our team can further analyze the businesses that opened and closed in the focus areas along the bike lanes and the non-bike lane area.

### 6.5 Geo Map (Bike Lane/Non-Bike Lane)

Utilizing the Polygon Filter setting in the Geo Map, the new and closed businesses can be isolated so a more focused analysis can be developed along the bike lane / non-bike lane area.

From 2016 to 2021 along the Bike Lane: 139 new businesses opened, 82 closed, and 17 businesses opened and closed during the same period. The closure rate for businesses that opened and closed during the same period was 12%. Along Santa Monica Blvd. (Non-Bike Lane) 94 new businesses opened, 92 closed, and 23 opened and closed during the same period. The closure rate along Santa Monica Blvd. was 24%.

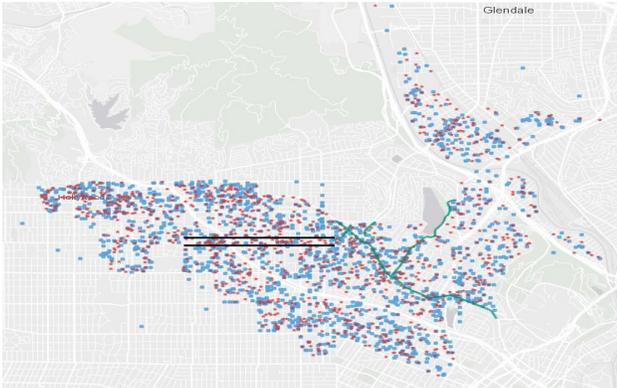

Figure 8. Area of study Geo Map with open/closed.

These results confirm our hypothesis that businesses near established bike lane infrastructure are more resilient to closures compared to businesses without the bike lane infrastructure. The results of this small sample size suggests the study warrants a more thorough review of the data and expansion to focus on all bike lane infrastructure throughout Los Angeles.

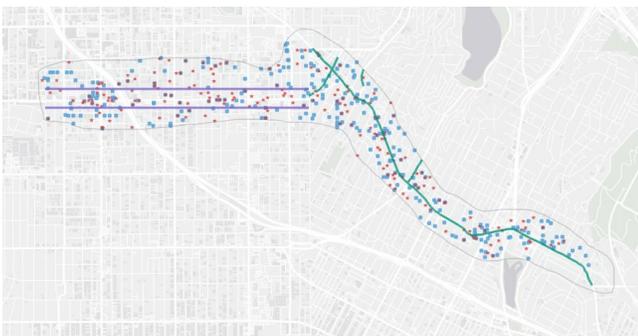

Figure 9. Focused area of study Geo Map.

## 7. Conclusion

Finally, we can conclude the following:
- More new business opened on the bike lane during the period of study than the non-bike lane route.
- The same amount of business closed during the period of study on both the bike lane and non-bike lane route.
- New businesses along the bike lane route were less likely to close. (12% vs 24%)
- Bike lanes appear to positively affect local business resilience.
- Further research is needed in order to conclude positive economic benefit of bike lanes.

Furthermore, our analysis shows that COVID-19 may not have had as significant of an impact in closures then predicted. There is no clue to tease out the actual impact of COVID-19. But the data shows that bike lanes themselves played a significant role in predicting which business remained open.

For the future work, we need further analysis with a larger dataset to give insights on bike lane infrastructures and their supporting/insulating of local businesses from closure and the potential positive economic impact.